\documentclass[journal,twocolumn]{IEEEtran}
\usepackage[round]{natbib}
\usepackage[pdftex]{graphicx}
\usepackage{amsmath}
\usepackage{ulem}
\usepackage[justification=centering, font=small]{caption}
\usepackage{cuted}
\usepackage{subcaption}
\usepackage[bottom]{footmisc}
\renewcommand{\footnoterule}{\vfill\kern -3pt \hrule width 0.4\columnwidth \kern 2.6pt}

\ifCLASSINFOpdf
\else
\fi
\usepackage{xcolor}

\graphicspath{{figures}}

\begin{document}
%
\title{Assessing the Time Evolution of COVID-19 Effective Reproduction Number in Brazil}
%
%
%

\author{Edson Porto da Silva,~\IEEEmembership{Senior~Member,~IEEE,}
        and~Antonio~M.N.~Lima,~\IEEEmembership{Senior~Member,~IEEE}
\thanks{Edson Porto da Silva and Antonio~M.N.~Lima are with the Department
of Electrical and Computer Engineering, Center for Electrical Engineering and Informatics, Universidade Federal de Campina Grande).}
}

\markboth{Anais da Academia Brasileira de Ciências,~Vol.~XX, No.~X, Novembro~2022}%
{Shell \MakeLowercase{\textit{et al.}}: Bare Demo of IEEEtran.cls for IEEE Journals}
\maketitle


\begin{abstract}
In this paper, we use a Bayesian method to estimate the effective reproduction number ($R(t)$), in the context of monitoring the time evolution of the COVID-19 pandemic in Brazil at different geographic levels. The focus of this study is to investigate the similarities between the trends in the evolution of such indicators at different subnational levels with the trends observed nationally. The underlying question addressed is whether national surveillance of such variables is enough to provide a picture of the epidemic evolution in the country or if it may hide important localized trends. This is particularly relevant in the scenario where health authorities use information obtained from such indicators in the design of non-pharmaceutical intervention policies to control the epidemic. A comparison between $R(t)$ estimates and the moving average (MA) of daily reported infections is also presented, which is another commonly monitored variable. The analysis carried out in this paper is based on the data of confirmed infected cases provided by a public repository. The correlations between the time series of $R(t)$ and MA in different geographic levels are assessed. Comparing national with subnational trends, higher degrees of correlation are found for the time series of $R(t)$ estimates, compared to the MA time series. Nevertheless, differences between national and subnational trends are observed for both indicators, suggesting that local epidemiological surveillance would be more suitable as an input to the design of non-pharmaceutical intervention policies in Brazil, particularly for the least populated states.
\end{abstract}

\IEEEpeerreviewmaketitle

\section{Introduction}

\IEEEPARstart{T}{HE} spread of the SARS-CoV-2 which led to the COVID-19 pandemic has brought many challenges to policymakers around the globe. To prevent explosive growth in the number of infections that may result in the collapse of the health systems, non-pharmaceutical interventions (NPI) had to be put in place to contain the virus spread. NPI is any non-pharmaceutical measure aimed to reduce the number of contagious contacts in the population, from promoting the use of hand sanitizers to lockdowns. However, NPIs such as closing schools and non-essential services, or implementing full lockdowns have significant negative economic side-effects, making them difficult to be sustained over long periods. Hence, the health authorities faced the hard task of defining epidemiological policies, trying to find the best balance in adopting the necessary NPIs to contain the virus spread, while minimizing the undesired consequences to the population. Defining the appropriate policy requires a careful assessment of the pandemic status and its evolution. In this context, the use of mathematical epidemiological models is paramount \citep{Kermack1927,Brauer2008, Britton2019b}. Unsurprisingly, since the beginning of the COVID-19 pandemic in 2020, many studies have been published exploring different categories of models to assess the status of the number of infections and/or to forecast its evolution in a given population \citep{Saldana2021}. 

The availability of daily updated data sets on, e.g., the reported number of infected cases and the reported number of deaths due to COVID-19 allowed data-driven models to unveil several aspects of the pandemic. Nevertheless, two different models designed to assess similar epidemiological variables may be built based on distinct sets of assumptions. Such conceptual differences will most certainly affect the interpretations and conclusions drawn from the outcome of each model. That is particularly important to quantify the underlying uncertainty associated with the forecast of each model \citep{Hutson2020, Thompson2020, Gostic2020}. The decision-making process guided by information provided by any epidemiological model will always be subject to uncertainty and risk. Hence, proper scrutiny is always necessary before using outcomes from a given model, e.g., to support changes in NPIs or to convey information to the public \citep{Thompson2020}.

The deterministic susceptible-infected-recovered (SIR) compartmental model \citep{Kermack1927} and its variations were widely explored during the COVID-19 pandemic. These models are based on a few simplifying assumptions about the dynamics of the disease spreading in a population and are often more \textit{user friendly} to estimate, e.g., peaks of infections, the percentage of the population that will get infected in scenarios with or without the adoption of NPI measures \citep{Morris2021}. Unfortunately, many sources of uncertainty affect the parameter estimation for those models, limiting their ability to forecast the epidemic evolution. When more realistic assumptions are introduced, the complexity of such models can quickly scale up, and extracting reliable information from them also becomes cumbersome, mainly due to a large number of parameters and the limited amount of available data. Uncertainty related to the transition process between compartments can be managed, to some extent, by stochastic epidemic models \citep{Britton2019b}.

Rather than attempting to predict absolute figures in the number of infections or deaths, SIR-type models and other epidemiological models  can be used to infer the \textit{relative outbreak potential} of the infections over time by estimating the so-called \textit{effective reproduction number} $R(t)$, which provides an important picture of the epidemic evolution over time. However, estimates of $R(t)$ also need to be considered with care, especially because the real-world data available for the estimation process may provide an incomplete picture of the reality \citep{Britton2019,Gostic2020}. Random reporting delays, the backlog of reported infection data, a large number of unreported asymptomatic infections, and a lack of an effective testing policy are all examples of real-life issues that may influence the likelihood of the data reflecting the underlying reality. Additionally, in each country, the epidemic evolution is likely to be dependent on sub-national characteristics, like demographics, and inter-regional migration dynamics. 

Considering the many sources of randomness that may influence epidemiological data, when using information from the surveillance of epidemiological parameters to define NPIs, a proper assessment of uncertainty is necessary \citep{Morgan2021}. In that sense, probabilistic Bayesian models, which naturally are able to handle uncertainty, have an advantage. On the other hand, NPI policies can be defined based on no specific epidemiological model, but rather direct statistics obtained from the data on reported infections, such as the evolution of its moving average (MA) over time. In fact, during the COVID-19 pandemic, the weekly MA of reported infections was widely reported by the media to picture the evolution of the pandemic.

In this paper, we present an analysis of the $R(t)$ estimates to characterize the time evolution COVID-19 pandemic in Brazil. The investigation focuses on evaluating the degree of similarity in the development of the pandemic over time across the country by comparing national trends to the trends observed at sub-national levels, such as geographic regions, and states. We highlight the similarities and differences among estimates made with data collected at different geographic levels: country, region, and state. The correlations between national and sub-national $R(t)$ estimates are discussed, as well as their implications when considering the design of NPI policies.

\section{Epidemic reproduction numbers}

In this section, the fundamentals of epidemiological reproduction numbers are briefly reviewed and a mathematical model for the effective reproduction number $R(t)$ is discussed. Moreover, the Bayesian framework considered in this work to estimate $R(t)$ from data on reported infections is also presented.


\subsection{Basic reproduction number versus effective reproduction number}

The basic reproduction number $R_0$ (R-naught) is arguably the most important parameter to evaluate the potential for an infectious disease to cause large epidemic outbreaks \citep{Britton2019}. It is defined as the average number of secondary infections an infected individual will generate in a large population at the beginning of an infectious-disease outbreak, where approximately all individuals are susceptible.  Consequently, if $R_0<1$, the disease will not have the potential to spread over a significant fraction of the population, while if $R_0>1$ a large outbreak will almost certainly occur. Therefore, for a given susceptible population, the $R_0$ value of an infectious disease is a fingerprint of its epidemic potential. The assumption of \textit{homogeneous mixing} is usually applied to define $R_0$ in a few scenarios, meaning that any infected individual in the population has the same probability of transmitting the disease to any susceptible individual.

The importance of $R_0$ relates to the fact that it is a measure of the potential for a disease to spread over a large susceptible population. However, the assumptions used to derive $R_0$ may reflect well the conditions at the beginning of an epidemic. Still, they tend to become less realistic as the spreading of the disease evolves. Two main reasons corroborate this fact. First, once an infectious-disease outbreak starts, non-pharmaceutical or pharmaceutical measures can take place to control the disease spreading. Consequently, the chances of a susceptible individual becoming infected will change over time, e.g., depending on the severity of the NPI measures. Second, even in the absence of NPI measures, complex dynamics linked e.g. to the social structure in which the individuals are inserted will influence the epidemic evolution. Therefore, knowing $R_0$ is insufficient to guide the health authorities' decisions in managing NPIs as the epidemiological conditions change over time. Instead, the instantaneous effective reproduction number $R(t)$ \citep{Fraser2007a, Cori2013, Flaxman2020, Gostic2020} provides a better picture of an ongoing epidemic. 

The instantaneous effective reproduction number captures, to some extent, the time-varying behavior of the disease spreading. Differently from $R_0$, the definition of $R(t)$ does not assume that the entire population is susceptible, i.e. it can be applied to any moment in time where an unknown fraction of the population is can be infected. The value of $R(t)$ can be interpreted as the average number of secondary infections that an infected individual is expected to produce should the conditions at time $t$ remain unchanged \citep{Fraser2007a}. When $R(t)>1$, the number of newly reported infections is likely to increase, whereas $R(t)<1$ suggests it is likely to decrease. If $R(t)=1$, the number of new infections is likely to remain stable. Moreover, the value of $R(t)$ is also related to how fast the number of infections is increasing or decreasing. The main advantages of this approach are the capability of $R(t)$ to track the time-varying epidemic state and the fact that its estimation relies only on statistics from the number of reported infections and a few specific characteristics of the disease \citep{Cori2013,Bettencourt2008}, such as the generation time distribution or the serial interval \citep{Britton2019b}. Hence, the effective reproduction number provides useful information to assist decisions by the health authorities during epidemic outbreaks.

\begin{figure}[h!]
    \centering
    \includegraphics[width=0.85\linewidth]{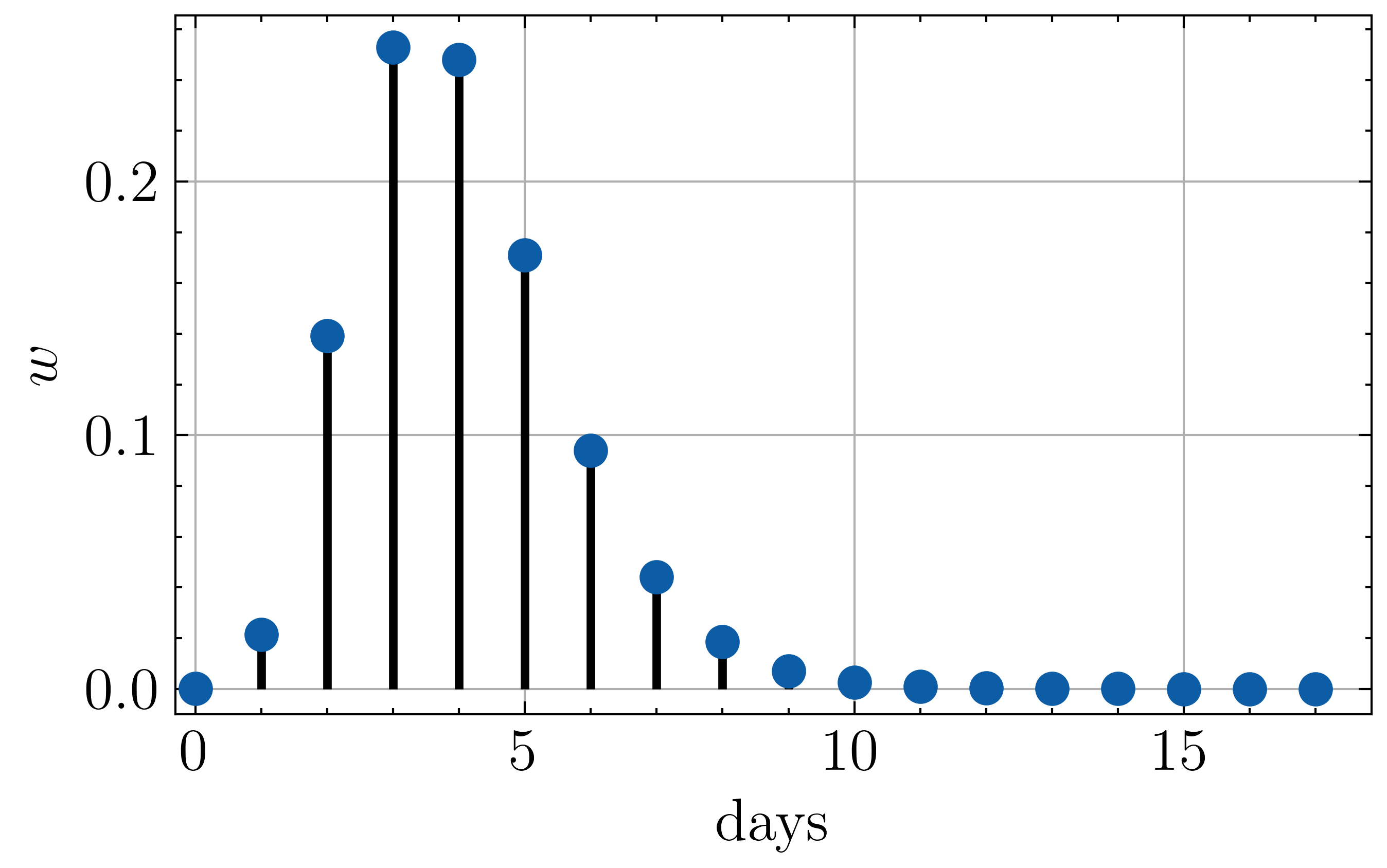}
    \caption{\footnotesize{Serial interval distribution/generation time distribution $\left\lbrace w_j \right\rbrace_{j=1}^{n}$ assumed for COVID-19 \citep{Flaxman2020}.}}
    \label{genTimeDist}
\end{figure}

Following \citep{Fraser2007a}, $R(t)$ can be defined as 

\begin{equation}\label{Rt_eq_1}
R(t)=\frac{I(t)}{\int_{0}^{\infty} I(t-\tau) w(\tau) d \tau}
\end{equation}

\noindent where $I(t)$ is the number of infected individuals at time $t$ (calendar time), and $w(\tau)$ is the generation time distribution, i.e., the distribution of the new infections as a function of time of infection.
A discrete-time version of (\ref{Rt_eq_1}) is given by
\begin{equation}\label{Rt_eq_2}
R\left(t_{i}\right)=\frac{I_{i}}{\sum_{j=1}^{n} w_{j} I_{i-j}}
\end{equation}

\noindent where $ \left\lbrace w_j \right\rbrace_{j=1}^{n}$ is a discrete-time version of $w(\tau)$. It is worth mentioning that the generation time distribution is rarely known since the times of infection cannot be observed in practice. Instead, the serial interval distribution, which corresponds to the distribution of time intervals for symptoms onset between the infected and the infectee is used as an approximation. It is important to emphasize that uncertainties associated with the generation time distribution will affect the uncertainty in estimating $R(t)$ over time. For instance, different variants of SARS-CoV-2 may result in distinct generation time distributions and serial intervals for COVID-19 \citep{Hart2022,Pung2021}. Nevertheless, in the context of monitoring the spreading of a new disease, it is realistic to assume that information about the generation time distribution of the disease is limited. Here, as in \citep{Flaxman2020}, the generation time distribution of COVID-19 is approximated by its serial interval distribution, shown in Fig.~\ref{genTimeDist}.

The definition in (\ref{Rt_eq_2}) provides a direct expression to estimate $R(t_i)$ from the data on reported infections $I_i$. The interpretation of $R(t_i)$  becomes evident after simply rewriting (\ref{Rt_eq_2}) as

\begin{equation}\label{Rt_eq_3}
I_{i} = R\left(t_{i}\right)\sum_{j=1}^{n} w_{j} I_{i-j}.
\end{equation}

\noindent Thus, from (\ref{Rt_eq_3}), the expected number of new infections $I_{i}$ at day $i$ is given by the product of the effective reproduction number $R(t_i)$ and the total infectivity at day $i$. The total infectivity is the sum of the number of new infections reported in the past days weighted by the generation time distribution. The interval of past days considered in the sum depends on the length $n$ of $ \left\lbrace w_j \right\rbrace_{j=1}^{n}$. To simplify notation, hereafter we refer to $\left\lbrace w_j \right\rbrace_{j=1}^{n}$ only as $\mathbf{w}$. Moreover, we will use $I_{t_{1}:t_{2}}$ to denote $I_{t_{1}},I_{t_{1}+1},\dots,I_{t_{2}}$, $t_{1}\leq\ t_{2}$.

In epidemiological modeling, reproduction numbers can also be directly associated with expressions involving parameters of deterministic or stochastic compartmental models, e.g., the infection and recovery rates used in the SIR model \citep{Britton2019b}. Thus, when fitting models with epidemiological data, reproduction numbers can also be inferred from the fitted parameters of each model. However, parameter estimation for such models can be challenging, since it may require data that is not easily available, other than just the number of reported infections (i.e., data on the number of susceptible individuals, number of recovered individuals, etc.).

In general, the data collected during an epidemic outbreak is incomplete. Particularly in the case of the COVID-19 pandemic, the number of daily reported infections or deaths is commonly the only available information that can be used to infer epidemiological parameters, such as reproduction numbers. Other sources of data, such as the number of hospitalizations and the positive test rates may also be available, depending on the country. During the COVID-19 pandemic in Brazil, due to the lack of a standard policy for random testing of a large and diverse population, as well as the absence of a unified system to collect hospitalizations data all over the country, this information was either unavailable or difficult to use (e.g. due to high variability). Under such conditions, a model that relies only on daily reported infections (\ref{Rt_eq_2}) becomes a useful tool to estimate the effective reproduction number.

\subsection{Estimating the effective reproduction number}\label{SecIIB}

In this paper, we consider the effective reproduction number estimator proposed in \citep{Cori2013}, which requires only information about the generation time distribution and the data on reported infections. This estimator is derived under the following assumptions:

\begin{enumerate}
    \item The generation time distribution is independent of the calendar time.
    \item Transmission of the disease is described as a Poisson process, i.e., the number of new infections at time $t_i$ is Poisson distributed with mean $m=R(t_i)\sum_{j=1}^{n}w_j I_{i-j}$. The likelihood of the number of new infections $I_i$, given $R(t_i)$ and the set of all previous observed number of infections $I_{0:i-1}$,
    is written as
    \begin{equation} 
        P\left(I_{i}|I_{0:i-1}, \mathbf{w}, R(t_i)\right)=\frac{\left(R(t_i) \Lambda_{i}\right)^{I_{i}} e^{-R(t_i) \Lambda_{i}}}{I_{i} !}
    \end{equation} 
    where $\Lambda_{i}=\sum_{j=1}^{n}w_{j} I_{i-j}$.
    \item $R(t)$ is constant over time periods $[i-\tau+1 ; i]$ of length $\tau$, which is indicated by the notation $R_{i,\tau}$. To simplify the inference process, a prior distribution for $R_{i, \tau}$ is chosen such that it is a conjugated prior for the Poisson distribution, i.e. a Gamma distribution.
\end{enumerate}

Given the assumptions above, the likelihood of the incidence $I_{i-\tau+1:i}$ during this time period given $w$, $R_{i,\tau}$ and $I_{0:i-\tau}$ is
\begin{equation}
  P\left(I_{i-\tau+1:i} | I_{0:i-\tau}, \mathbf{w}, R_{i, \tau}\right)=\!\!\!\!\!\!\prod_{j=i-\tau+1}^{i}\!\!\!\!\!\!\frac{\left(R_{i, \tau} \Lambda_{j}\right)^{I_{j}} e^{-R_{i, \tau}, \Lambda_{j}}}{I_{j} !}
\end{equation}
Adopting a Bayesian setting, by assuming a Gamma distributed prior for $R_{i, \tau}$ with parameters $(a,b)$, i.e., $R_{i, \tau}\sim \Gamma\left(a, b\right)$, the posterior joint distribution of $R_{i, \tau}$ is given in (\ref{posterior}).
\vspace{-0.5cm}
\begin{strip}
\begin{equation}
        \small
        P\left(I_{i-\tau+1:i}, R_{i, \tau}|I_{0:i-\tau}, w\right)=P\left(I_{i-\tau+1:i} | I_{0:i-\tau}, w, R_{i, \tau}\right) P\left(R_{i, \tau}\right)\label{posterior}
     \normalsize
\end{equation}
where 
\begin{equation}
        \small
        P\left(R_{i, \tau}\right)=\frac{R_{i, \tau}^{a-1} e^{-R_{i, \tau} / b}}{\Gamma(a)b^{a}},
     \normalsize
\end{equation}
$P\left(I_{i-\tau+1:i} | I_{0:i-\tau}, w, R_{i, \tau}\right)$ is given by (5) and thus
\begin{equation}
        \small
P\left(I_{i-\tau+1:i}, R_{i, \tau}|I_{0:i-\tau}, w\right)\propto R_{i, \tau}^{a + \sum_{j=i-\tau+1}^{i}I_j - 1 } e^{-R_{i, \tau}\left(\sum_{j=i-\tau+1}^{i} \Lambda_{j} + \frac{1}{b}\right)} \prod_{j=i-\tau+1}^{i} \frac{\Lambda_{j}^{I_{j}}}{I_{j}!}.
     \normalsize
\end{equation}
\end{strip}

Finally, it follows from (\ref{posterior}) that the posterior distribution of $R_{i, \tau}$ is also a Gamma distribution, given by
\begin{equation}
    R_{i, \tau}\sim \Gamma\left(a+\sum_{j=i-\tau+1}^{i}I_{j}, \frac{1}{\frac{1}{b}+\sum_{j=i-\tau+1}^{i}\Lambda_{j}}\right)
\end{equation}

The posterior mean $\mu$ and the posterior variance $\sigma^2$ of $R_{i, \tau}$ can be calculated as
\begin{align}
\small
    \mu &= \frac{a+\sum_{j=i-\tau+1}^{i} I_{j}}{\frac{1}{b}+\sum_{j=i-\tau+1}^{i} \Lambda_{j}}\label{Rt_mean}\\
    \sigma^2 &= \frac{a+\sum_{j=i-\tau+1}^{i} I_{j}}{\left[\frac{1}{b}+\sum_{j=i-\tau+1}^{i}\label{Rt_var} \Lambda_{j}\right]^2}
\end{align}
\normalsize

Therefore, given a time series of reported infections, the expressions in (\ref{Rt_mean}) and (\ref{Rt_var}) can then be readily used to obtain estimates for $R_{i, \tau}$. It should be highlighted that, via numerical simulations, the authors in \citep{Cori2013} have verified that this estimator does not present significant sensibility to underreporting of infection cases when the rate of underreported cases is constant over time. This is an important feature to consider when applying this estimator to COVID-19 data since there is a large fraction of asymptomatic infections that do not get reported. Nevertheless, in this work, there is no attempt to address the problem of underreporting rates of COVID-19 data in Brazil, meaning that the reporting rates are assumed to remain constant over time \citep{Balthazar2021}. This assumption is also corroborated by the fact that during the pandemic the was no broad testing policy implemented by the Brazilian government aiming to evaluate e.g. the number of asymptomatic infections. 

\section{Assessing the reported data of the COVID-19 pandemic in Brazil}
Monitored epidemiological parameters are often taken into account by the health authorities when establishing NPI policies to control an epidemic. A few of those parameters, such as $R(t)$, can be monitored periodically, e.g. from the data on daily reported infections. Hence, it is important to understand the caveats and pitfalls of using such indicators to implement policy at the national and sub-national levels, particularly for large countries such as Brazil. In this section, we focus on analyzing data-based estimates of $R(t)$ in national and different sub-national levels of Brazil and discuss its possible implications for the way NPI policies are implemented. Three different scenarios that may be relevant for the implementation of NPI policies are considered:

\begin{enumerate}
    \item Surveillance of epidemiological parameters based on nationwide data.
    \item Surveillance of epidemiological parameters based on statewide data.
    \item Surveillance of epidemiological parameters based on national geographic regions.
\end{enumerate}

\subsection{Data analysis}
The data on reported COVID-19 infections is obtained from the public repository \citep{CotaCovid19br2020}, which aggregates the data reported by several state health authorities in different geographic scales. For the analysis, we consider the data on reported infections in the interval from 20 May 2020 to 20 May 2021. During this period, most of the Brazilian population was unvaccinated and a number of NPI measures were adopted to contain the transmission rate of the virus. Hence, data reported within this period should be mostly independent of the vaccination process. In the ideal scenario, the data on reported infections would not be affected by e.g. the underreporting of cases, delays from infection to report, and the damming of reported data. However, in reality, the available data is subject to many, if not all, of those variables at once. The results presented in this paper were obtained by analyzing the raw official datasets available with statistics from the COVID-19 pandemic in Brazil. No particular pre-processing was applied to e.g. fix or remove peaks of reported infections artificially generated from the damming of reported data. Thus, emulating the conditions of how the information would be available to the health authorities during the pandemic. The estimator model assumes only a fixed generation time distribution for the whole period analyzed. However, different virus variants may be linked to different generation time distributions \citep{Hart2022}.

The $R(t)$ values are calculated using the estimator discussed in Section~\ref{SecIIB}, assuming a weekly time window, i.e., $\tau~=~7$~days. The length of the time window is chosen to average weekly seasonality effects in the reporting. Wider windows (e.g. $\tau~=~15$~days) could be used for the same purpose at the expense of increasing the lag in the estimation process. In fact, it should be noted that, even though ideally $R(t)$ would track the dynamics of the epidemic in real-time, random or unknown delays in between infection and reporting will eventually introduce lags in the estimation process, such that estimates on $R(t)$ in practice would reflect the state of the epidemic some time in the past.

Whenever mentioned, the correlation coefficient $r_{x y}$ \citep{papoulis2002} used to evaluate correlation between time series $x$ and $y$ is defined as 
\begin{equation}\label{corrCoeffCalc}
r_{x y}=\frac{\sum_{k=1}^{N}\left(x_{k}-\bar{x}\right)\left(y_{k}-\bar{y}\right)}{\sqrt{\sum_{k=1}^{N}\left(x_{k}-\bar{x}\right)^{2}} \sqrt{\sum_{k=1}^{N}\left(y_{k}-\bar{y}\right)^{2}}}
\end{equation}
where $N$ is sample size, $x_{k}, y_{k}$ are the individual sample points indexed with $k$,  $\bar{x}=\frac{1}{N} \sum_{k=1}^{N} x_{k}$ is the sample mean of $x$, and analogously for $\bar{y}$. The times series were named according to the country name and standard Brazilian state abbreviation names, i.e., \(x,y\in\left\{\text{BR},\text{AC},\text{AL},\text{AP},\ldots,\text{SE},\text{SP},\text{TO}\right\}\). The correlation coefficient is a measure of linear dependence between two time series.

\begin{figure}[!ht]
    \centering
    \includegraphics[width=0.85\linewidth]{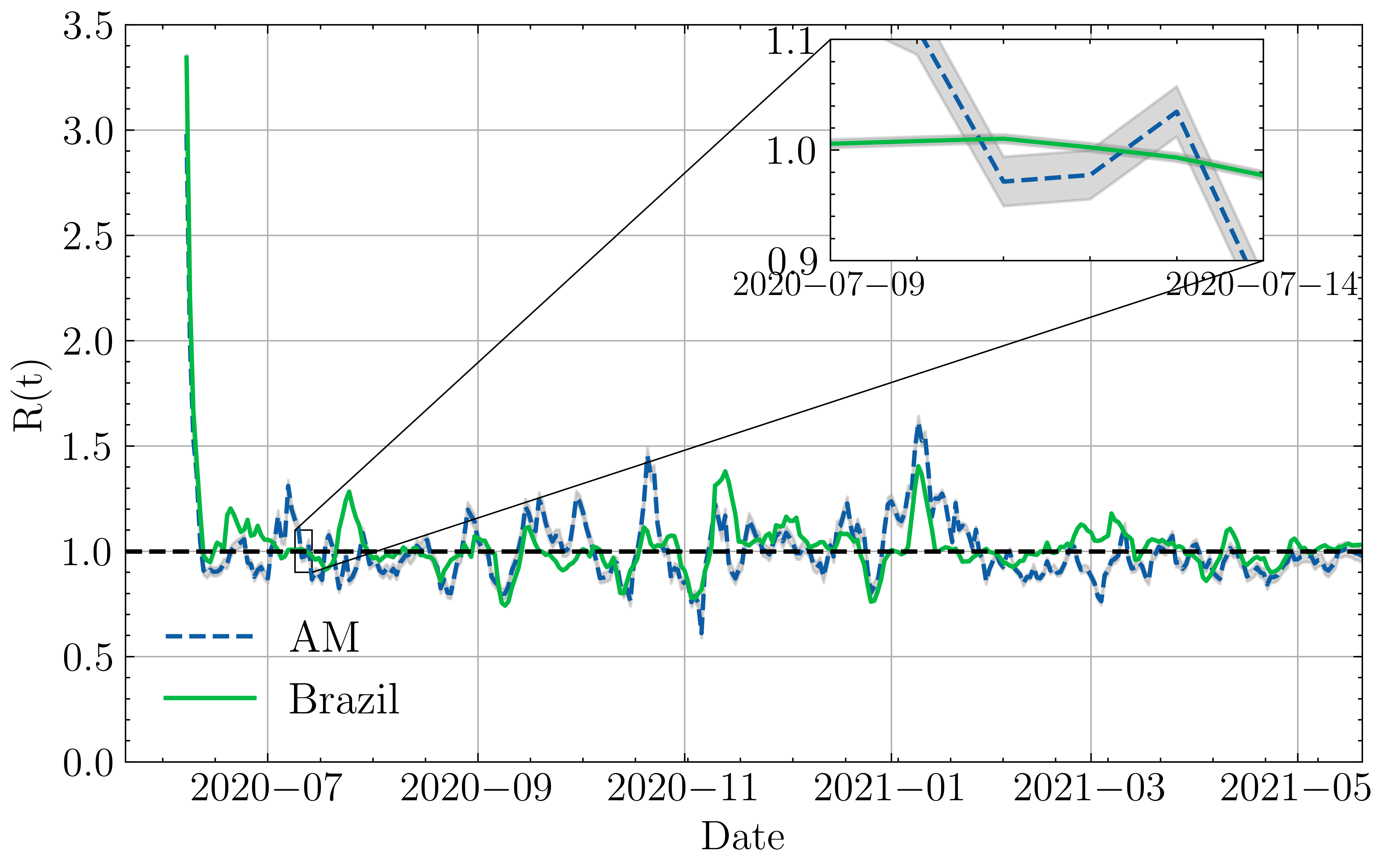}
    \caption{\footnotesize{$R(t)$ evolution in the estate of Amazonas (AM). Lines indicate the estimated mean of $R(t)$ and the shaded area delimits the $95\%$ credible interval.}}
    \label{RT_AM}
    \vspace{-0.25cm}
\end{figure}

\begin{figure}[!ht]
    \centering
    \includegraphics[width=0.85\linewidth]{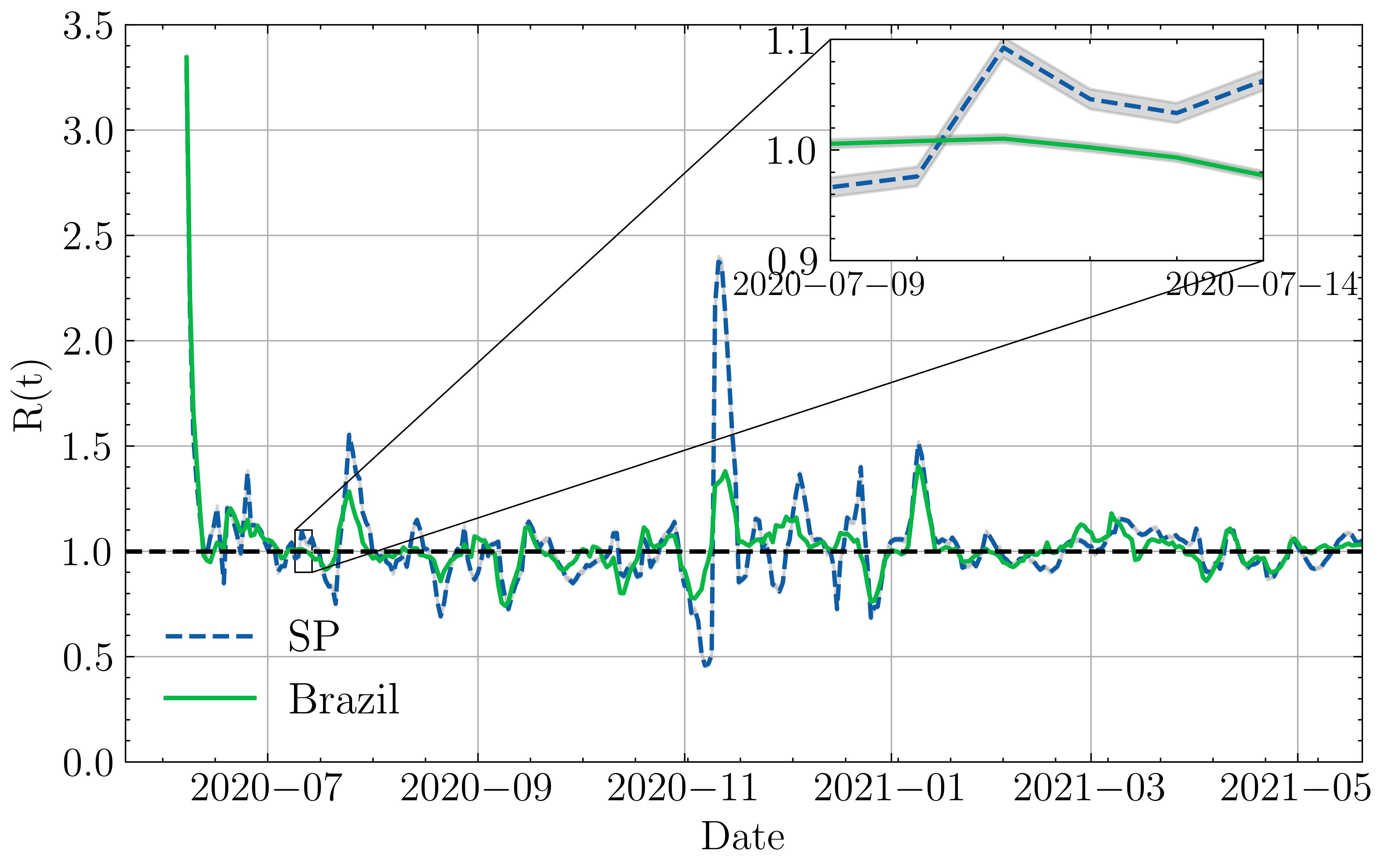}
    \caption{\footnotesize{$R(t)$ evolution in the estate of São Paulo (SP). Lines indicate the estimated mean of $R(t)$ and the shaded area delimits the $95\%$ credible interval.}}
    \label{RT_SP}
    \vspace{-0.25cm}
\end{figure}

\begin{figure}[!ht]
    \centering
    \includegraphics[width=0.85\linewidth]{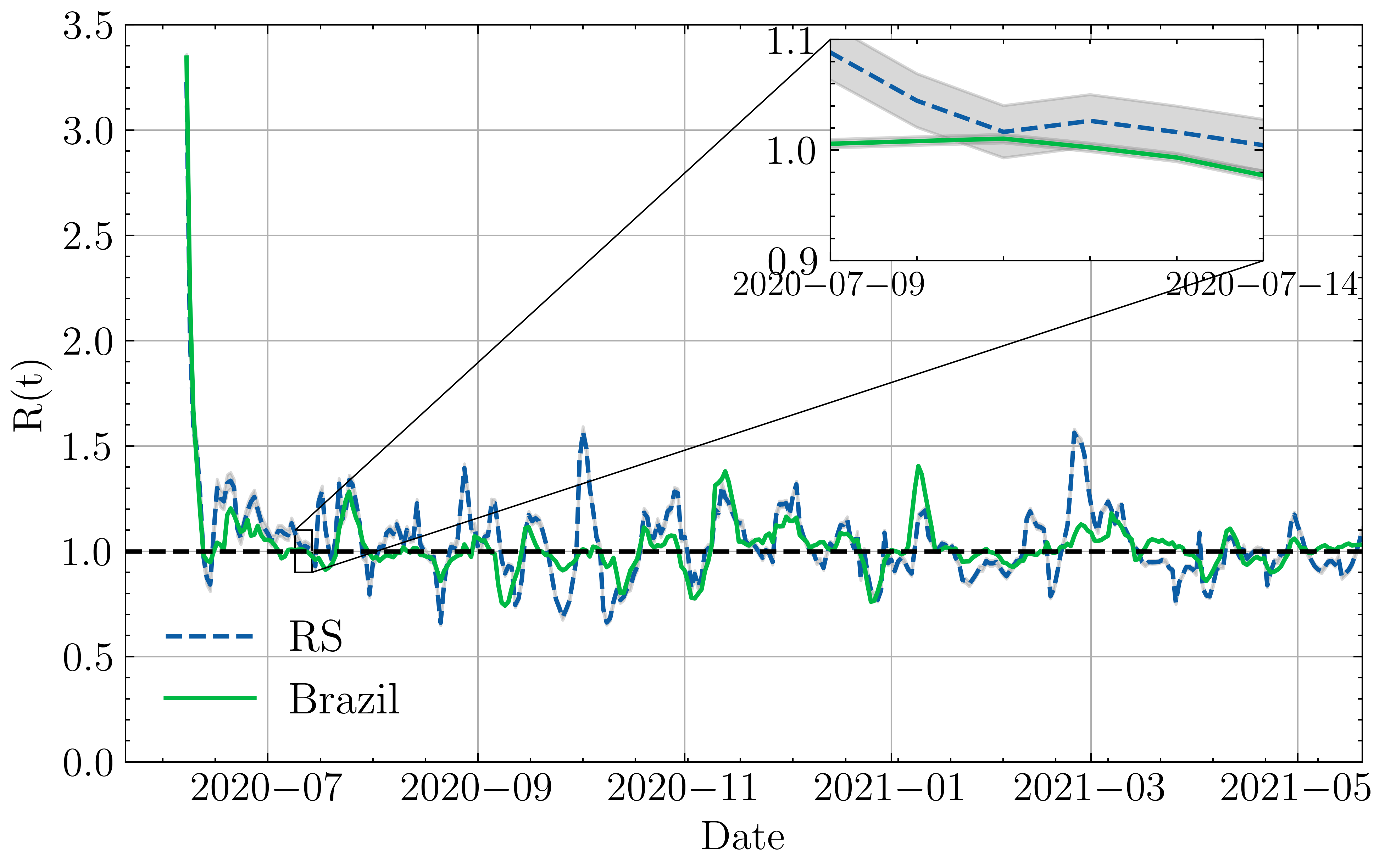}
    \caption{\footnotesize{$R(t)$ evolution in the estate of Rio Grande do Sul (RS). Lines indicate the estimated mean of $R(t)$ and the shaded area delimits the $95\%$ credible interval.}}
    \label{RT_RS}
    \vspace{-0.25cm}
\end{figure}

\begin{figure}[!ht]
    \centering
    \includegraphics[width=0.85\linewidth]{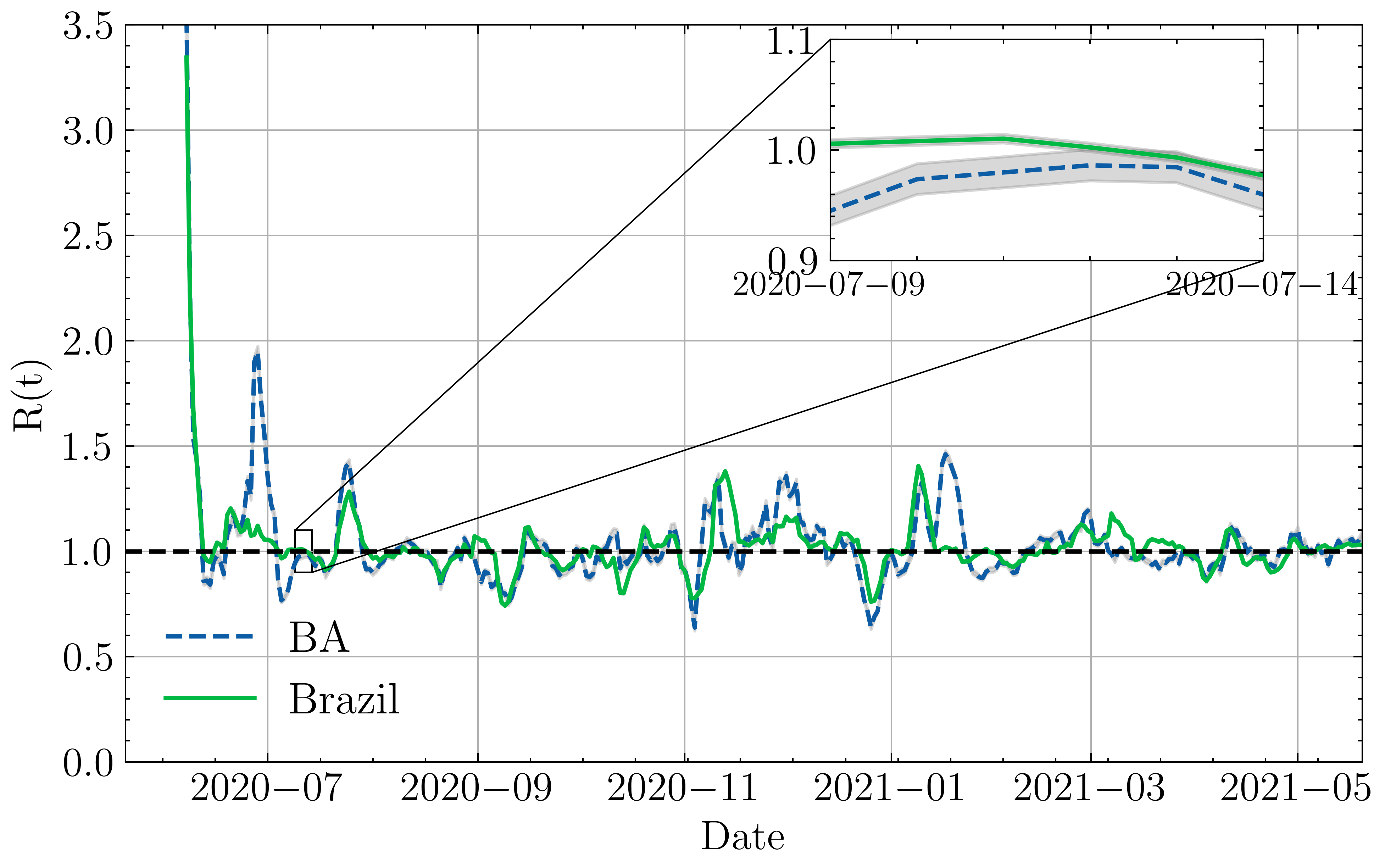}
    \caption{\footnotesize{$R(t)$ evolution in the estate of Bahia (BA). Lines indicate the estimated mean of $R(t)$ and the shaded area delimits the $95\%$ credible interval.}}
    \label{RT_BA}
    \vspace{-0.25cm}
\end{figure}

\begin{figure}[!ht]
    \centering
    \includegraphics[width=0.85\linewidth]{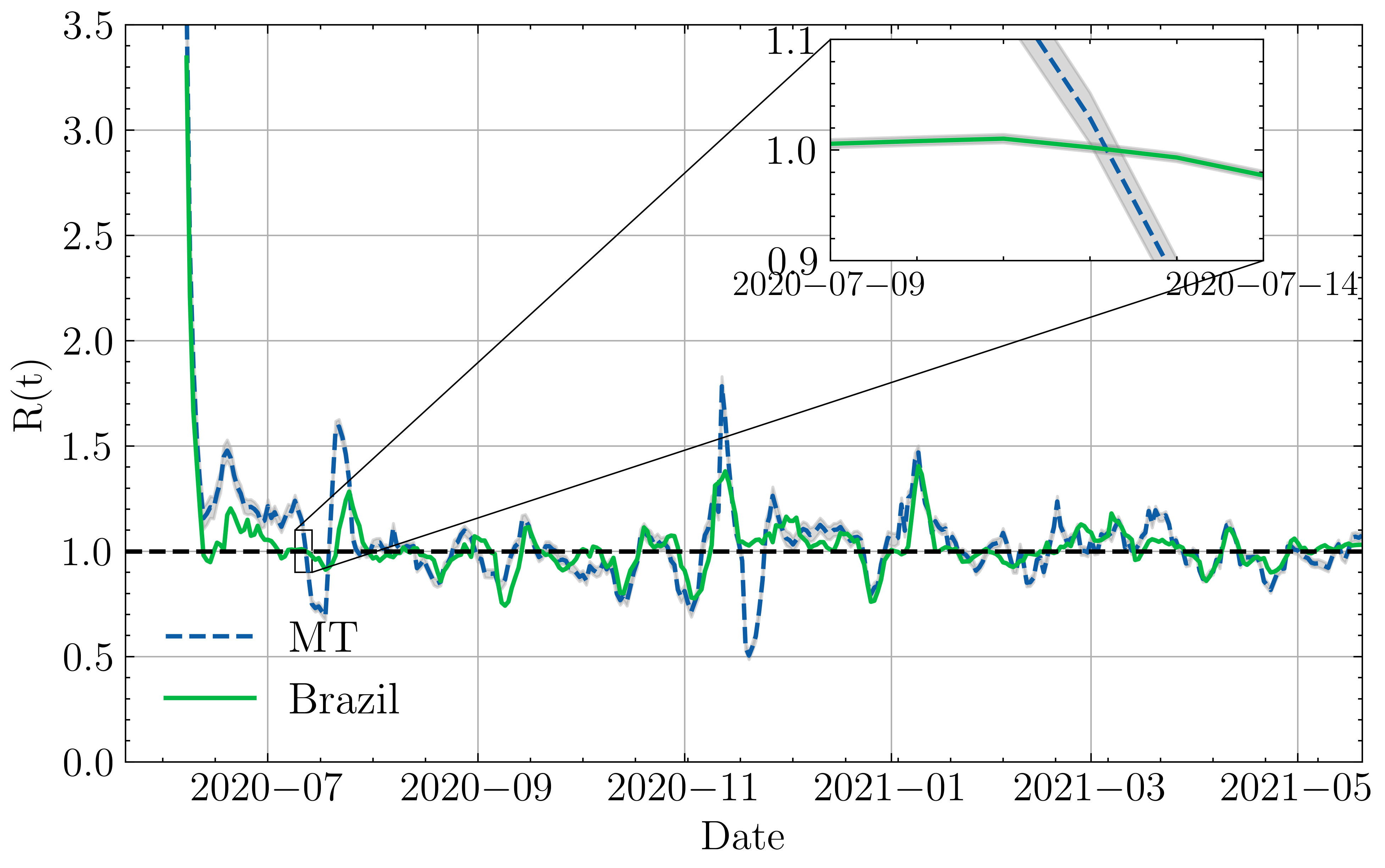}
    \caption{\footnotesize{$R(t)$ evolution in the estate of Mato Grosso (MT). Lines indicate the estimated mean of $R(t)$ and the shaded area delimits the $95\%$ credible interval.}}
    \label{RT_MT}
    \vspace{-0.25cm}
\end{figure}

In Fig.~\ref{RT_AM}-\ref{RT_MT}, the estimated mean of $R(t)$ (lines) and the $95\%$ credible interval (shaded area) as a function of the calendar time is depicted for the most populated state of each of the major geographic regions of Brazil. Additionally, the estimated mean of $R(t)$ for the entire country is plotted for comparison. Overall, the curves show similar trends, which indicates a degree of similarity between the $R(t)$ at the state and the country level. Nevertheless, in all five plots, significant discrepancies are observed in a few time intervals. Such differences could be a result of pandemic evolution scenarios varying over time from state to state and/or issues related to the reporting data policy (delays, damming, etc.) performed by each state. Note that, due to the large number of cases reported, the credible intervals become significantly narrow around the mean, as shown in the insets in each figure. The larger the population of the estate, the narrower the credible interval, as expected from the estimation model. One should be careful when interpreting the credible interval since it informs the range of uncertainty of the $R(t)$ estimates given the available data and the selected model. In other words, it does not reflect the uncertainty associated with the model itself. In particular, the model of \citep{Cori2013} does not account for uncertainty due to underreporting, reporting delay, and overdispersion in the case counts.

\begin{figure}[!ht]
    \centering
    \begin{subfigure}[t]{0.5\textwidth}
        \centering
        \includegraphics[width=0.9\linewidth]{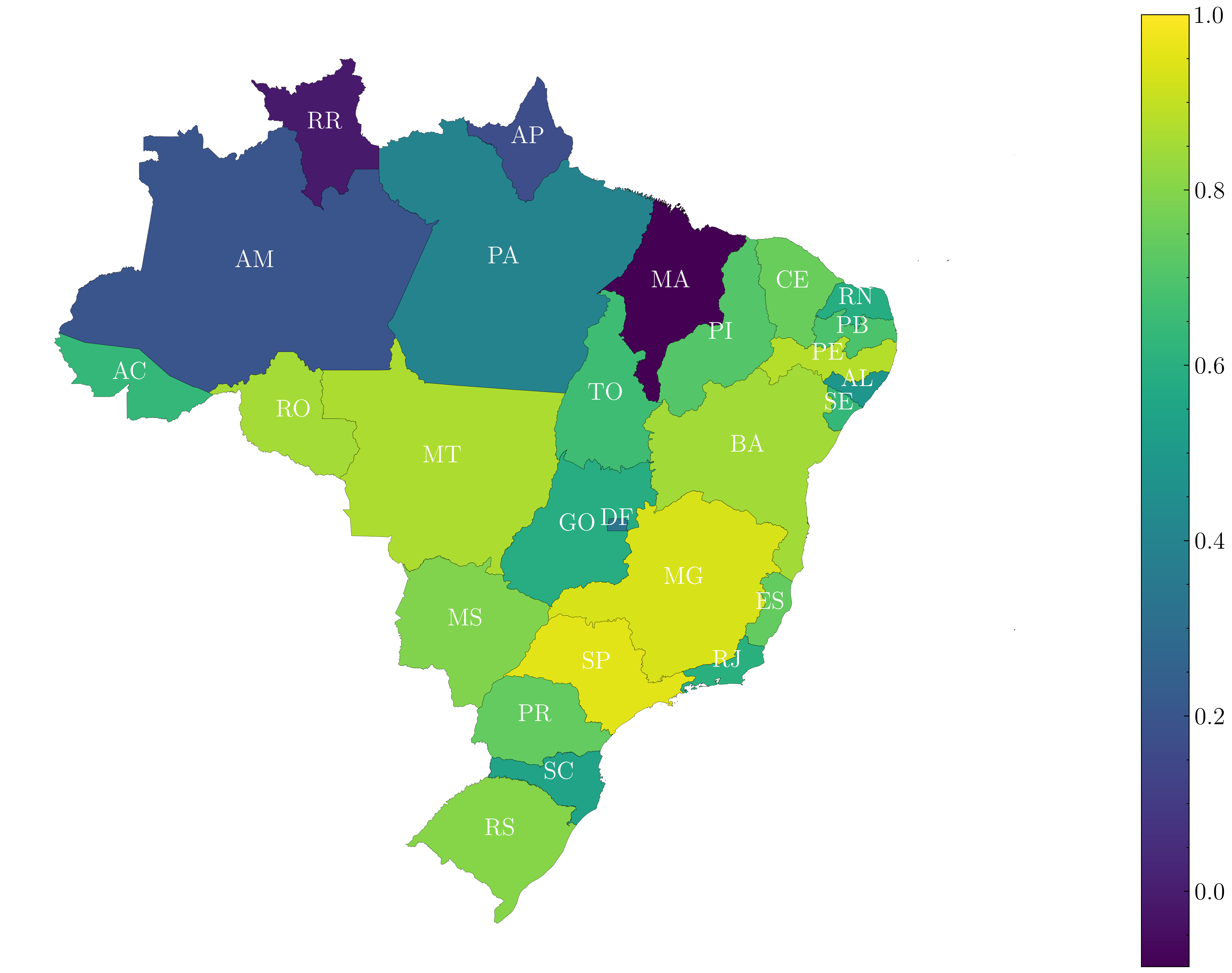}
        \caption{\footnotesize{Correlation coefficient between the moving average (7-day) of reported COVID-19 infection cases for each state with respect to the time series obtained by applying the same metric to the whole country.}}
    \label{heatMap_MA}
    \end{subfigure}%
    \\
    \begin{subfigure}[t]{0.5\textwidth}
        \centering
    \includegraphics[width=0.9\linewidth]{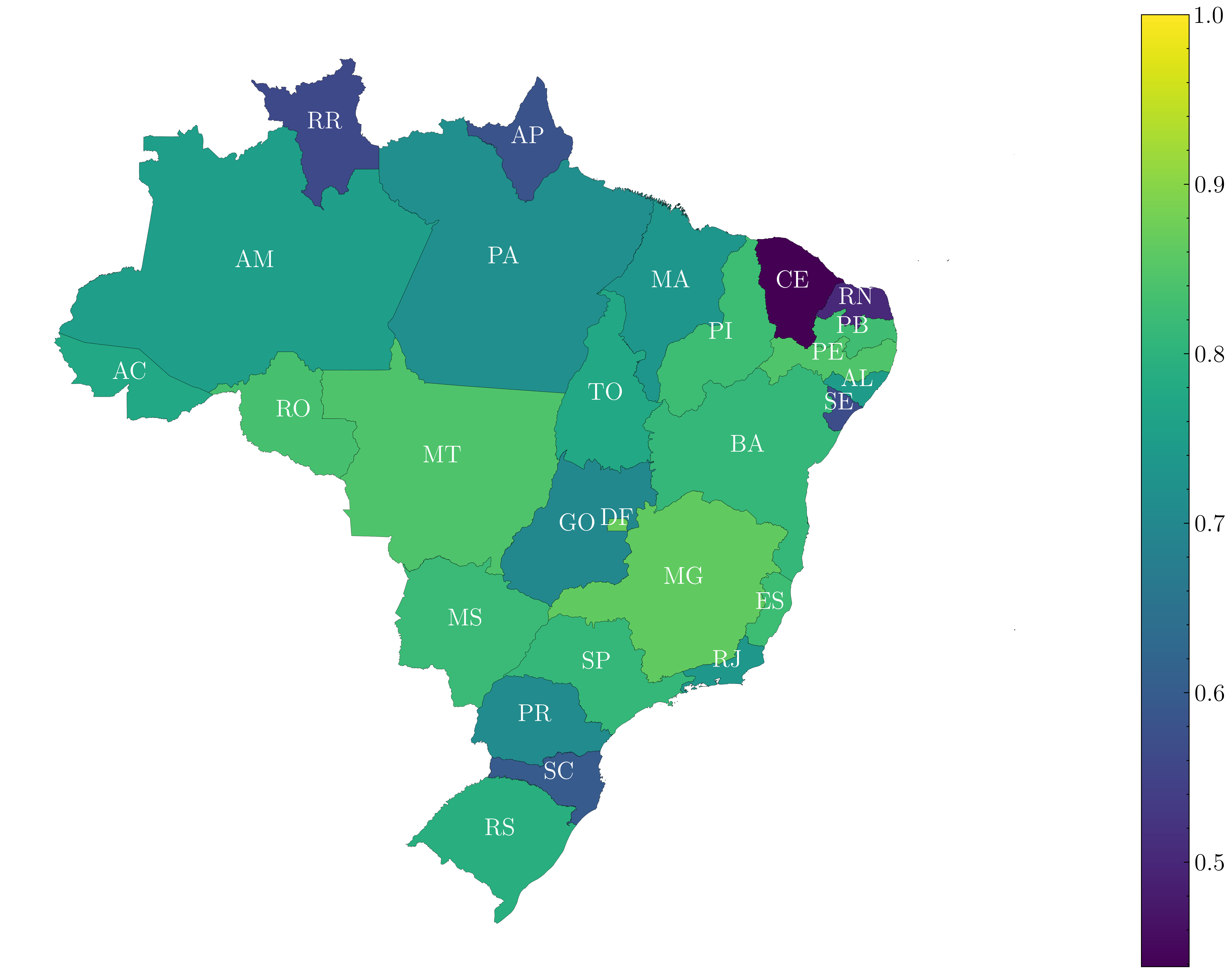}
    \caption{\footnotesize{Correlation coefficient between the mean estimated $R(t)$ for each state with respect to the time series obtained by applying the same metric to the whole country.}}
    \label{heatMap_Rt}
    \end{subfigure}
    \caption{\footnotesize{Heat map of the correlation coefficients of state versus country.}}
    \label{heatMap_all}
\end{figure}

\begin{figure*}[!ht]
    \centering
    \includegraphics[width=0.95\linewidth]{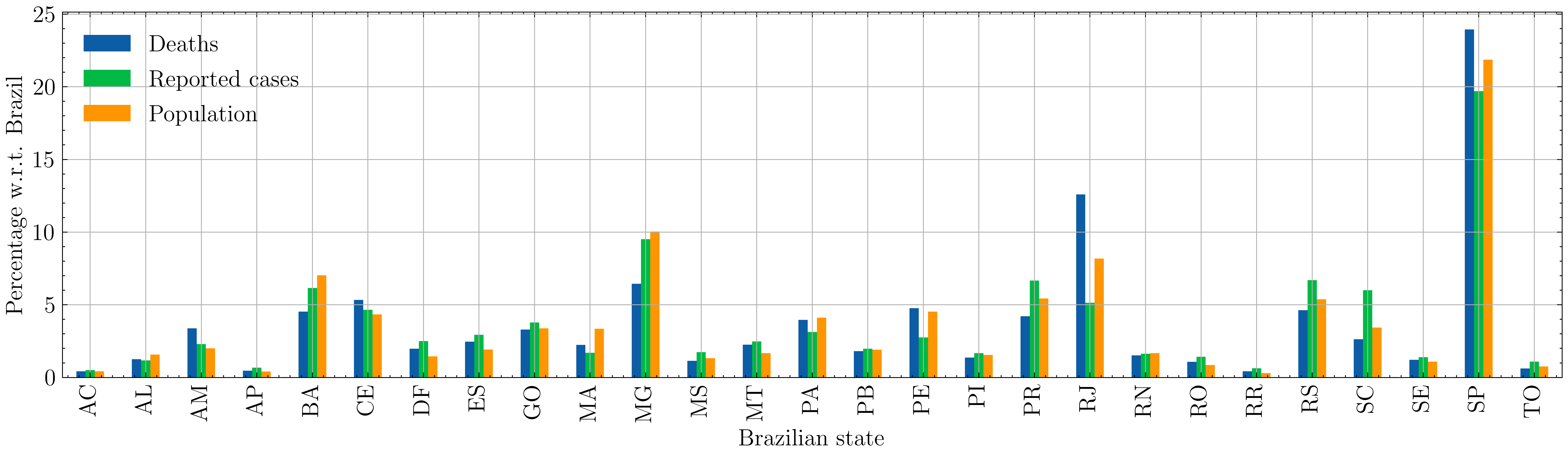}
    \caption{\footnotesize{Relative percentage of reported COVID-19 cases per state and relative population per state compared the overall data for Brazil.}}
    \label{relativeIncidence}
\end{figure*}

\begin{figure*}[!ht]
    \centering
    \includegraphics[width=0.95\linewidth]{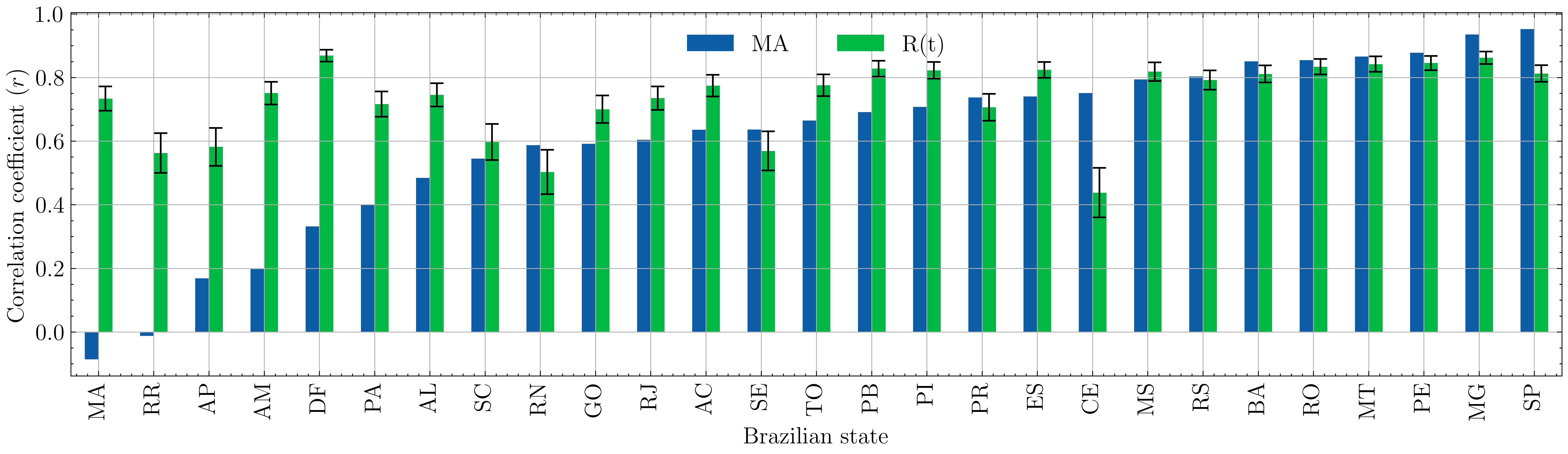}
    \caption{\footnotesize{Per state comparison of the correlation coefficients MA and $R(t)$ of each Brazilian state w.r.t the national trends. The error bars in the $R(t)$ correlation coefficients indicate the $95\%$ credible interval.}}
    \label{corrCoeff_all}
\end{figure*}

For comparison, the 7-day MA of reported cases is assessed. The MA time series is obtained by summing the number of reported infections in the last 7 days and dividing it by 7.
In Fig.~\ref{heatMap_MA}, a heat map is presented depicting the correlation coefficient of the 7-day MA curve of reported infections of each Brazilian state with the curve for the whole country. Overall, a relatively high correlation is observed between the time series of the country and of the states. For the interval analyzed, it is noted that the MA of reported infections in Brazil is highly correlated with the moving average of the states of São Paulo (SP) and Minas Gerais (MG). That can be linked to the fact that those are the two most populated Brazilian states, accounting for roughly 32\% of the Brazilian population. To better observe this trend, one can analyze the bar plots in Fig.~\ref{relativeIncidence}. The green bars represent the relative percent contribution of each state to the overall number of infections reported in the country. The orange bars show the percent fraction of the Brazilian population living in each state, according to data obtained from \citep{IBGE2022}. Assuming that every state is uniformly affected by the pandemic, it would be expected that the overall number of reported infections per state to be roughly proportional to its population. The results in Fig.~\ref{relativeIncidence} seem to support that hypothesis with a few exceptions. Notably, in the state of RJ, 5\% of the country reported infection cases were reported. However, in the same period, RJ has reported close to 13\% of the total number of COVID-19 deaths in Brazil.

In Fig.~\ref{heatMap_all} and in Fig.~\ref{corrCoeff_all}, a comparison of the correlation coefficients of the MA and the estimated mean of $R(t)$ is presented.
Credible intervals obtained by sampling time series from the posterior distributions of $R(t)$ were included in Fig.~\ref{corrCoeff_all} to account for the uncertainty in the calculation of the correlation coefficients between the times series of $R(t)$. Note that the states with wider credible intervals are mostly states with small populations (e.g. RR, AP, SE, RN), because those tend to be the ones with higher uncertainty in the $R(t)$ estimation. The widest credible interval is observed for CE, which is also the state with the lowest correlation with the national trend in terms of $R(t)$. On the other hand, there are also a few small states with high correlation and narrow credible intervals (e.g. DF, AL, AC, ES. RO). Hence, for those states, the evolution of $R(t)$ presented a higher degree of similarity with the overall national trend. In general, it would be always expected some degree of bias of the time series of MA and $R(t)$ at the national level towards the time series of the most populated regions. From this perspective, correlations between the time series of the monitored variables in the the least populated regions with the national trends are more informative to evaluate whether national surveillance will provide an effective description of the epidemic evolution in these areas. 

For the time period analyzed, the correlation of the MA data from states with the national data was clearly biased by the population of each state, whereas the correlation in the estimated $R(t)$ time series was less affected. In particular, for the states of MA, RR, AP, and AM, while $R(t)$ provides reasonable correlation levels, practically no correlation was observed in the moving average. Thus, when comparing national with state data, those results suggest that the MA of reported cases and the $R(t)$ estimates may provide different levels of agreement. Hence, such differences need to be considered when defining triggers for NPI policies based on those epidemiological surveillance indicators.

\subsection{Analysis per geographic region}
\begin{figure}[!ht]
    \centering
    \includegraphics[width=0.90\linewidth]{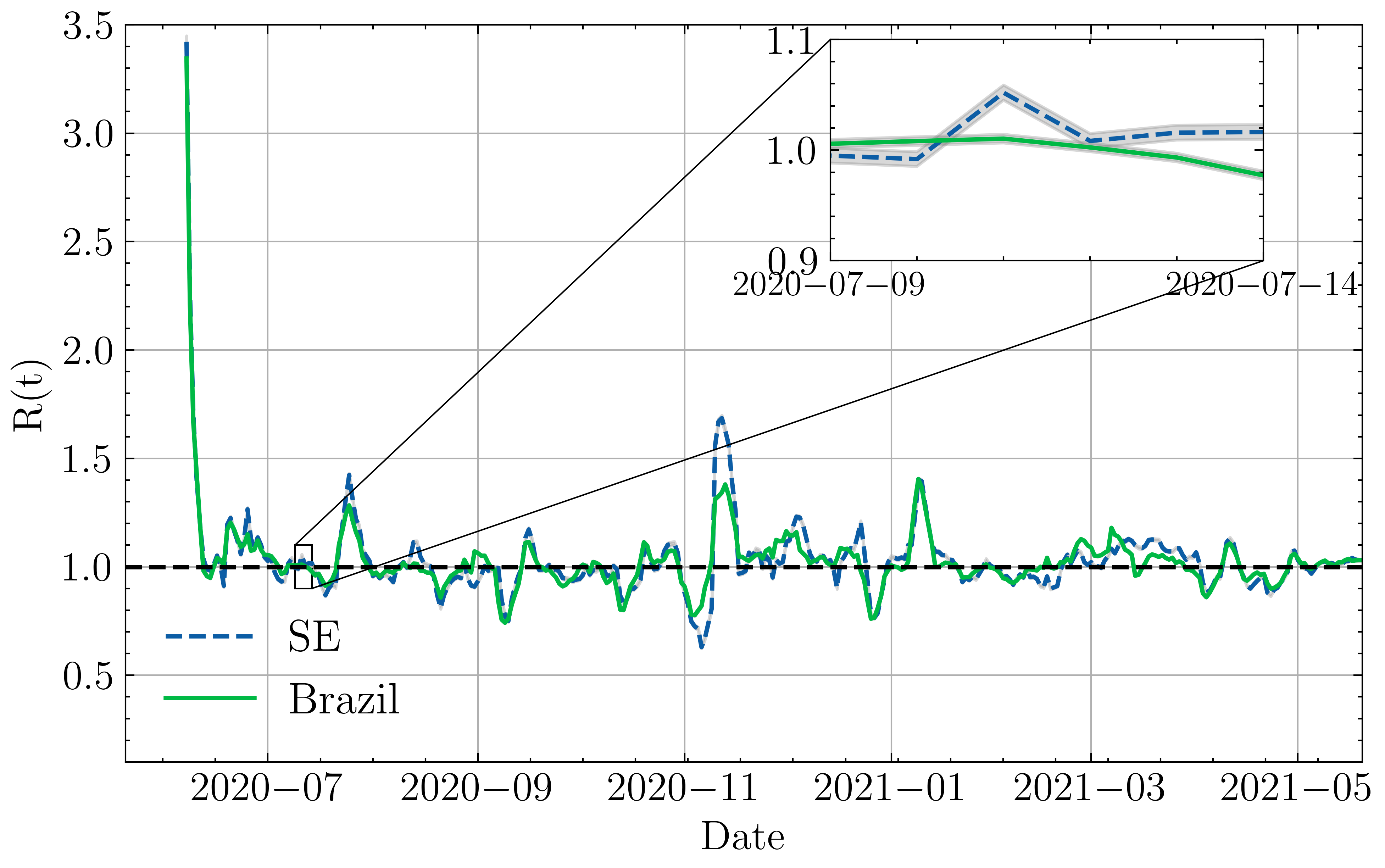}
    \caption{\footnotesize{$R(t)$ evolution in Southeast region (SE). Lines indicate the estimated mean of $R(t)$ and the shaded area delimits the $95\%$ credible interval.}}
    \label{RT_SE}
\end{figure}

\begin{figure}[!ht]
    \centering
    \includegraphics[width=0.90\linewidth]{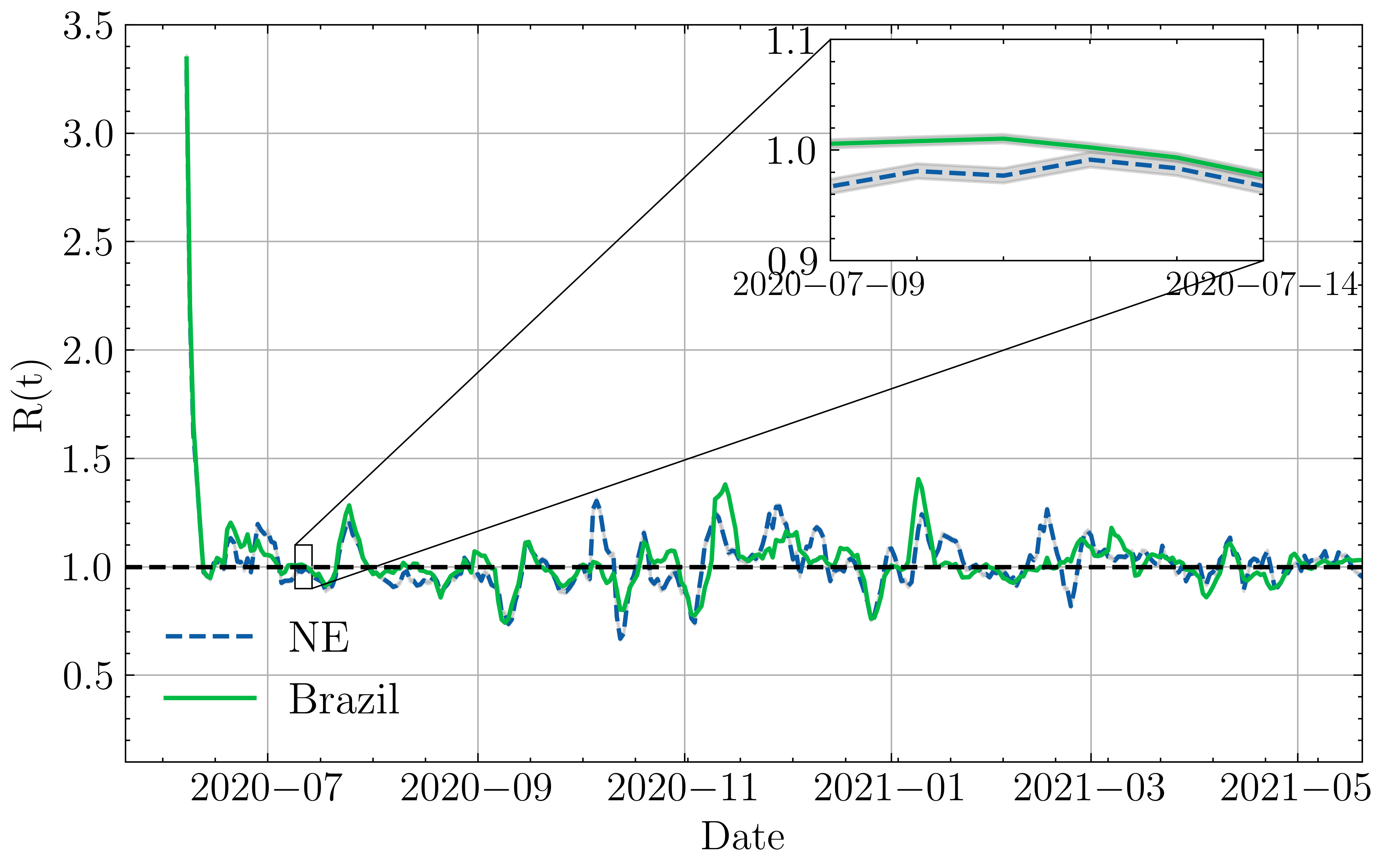}
    \caption{\footnotesize{$R(t)$ evolution in Northeast region (NE). Lines indicate the estimated mean of $R(t)$ and the shaded area delimits the $95\%$ credible interval.}}
    \label{RT_NE}
\end{figure}

\begin{figure}[!ht]
    \centering
    \includegraphics[width=0.90\linewidth]{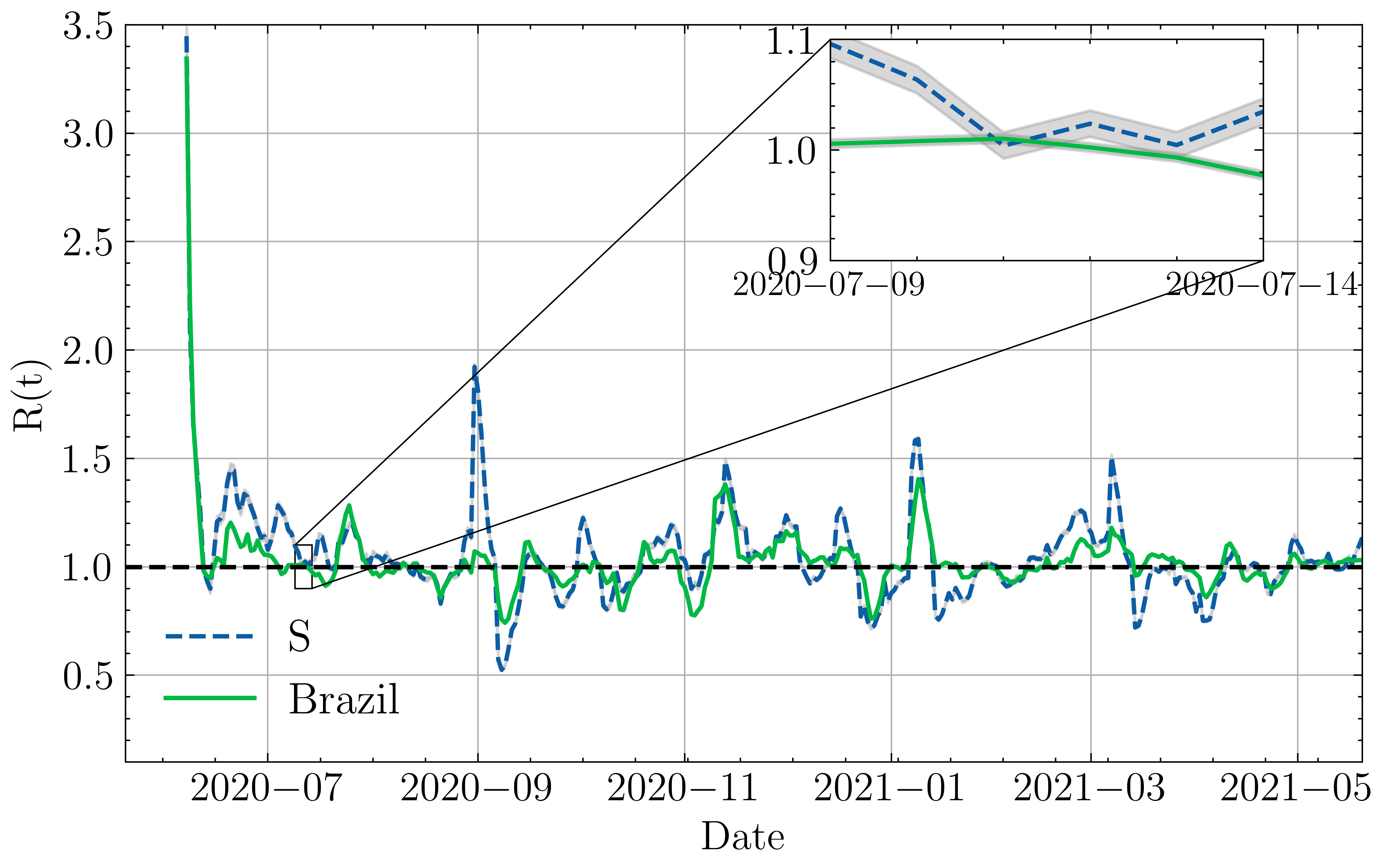}
    \caption{\footnotesize{$R(t)$ evolution in South region (S). Lines indicate the estimated mean of $R(t)$ and the shaded area delimits the $95\%$ credible interval.}}
    \label{RT_S}
\end{figure}

\begin{figure}[!ht]
    \centering
    \includegraphics[width=0.90\linewidth]{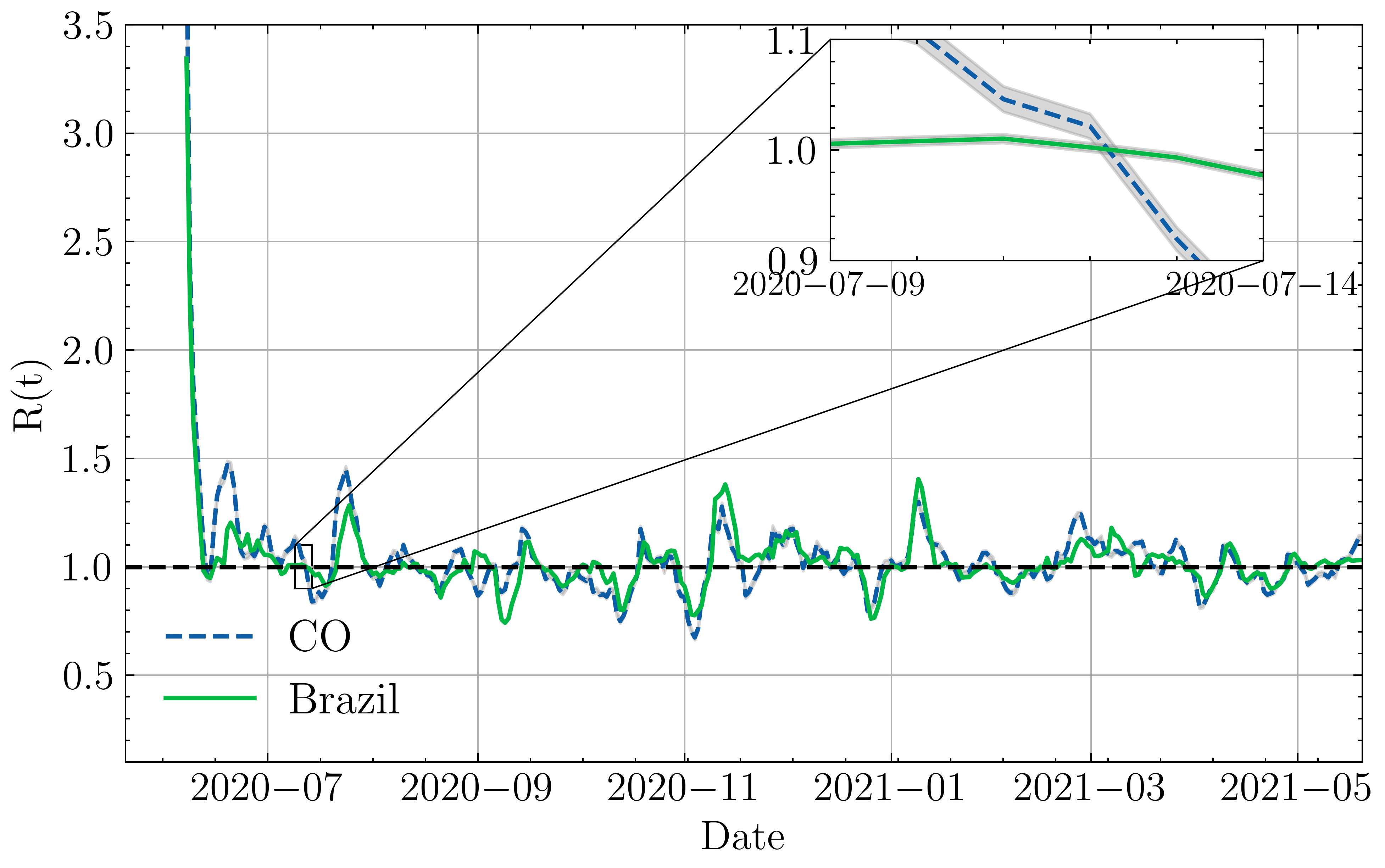}
    \caption{\footnotesize{$R(t)$ evolution in Center West region (CO). Lines indicate the estimated mean of $R(t)$ and the shaded area delimits the $95\%$ credible interval.}}
    \label{RT_CO}
\end{figure}

\begin{figure}[!ht]
    \centering
    \includegraphics[width=0.90\linewidth]{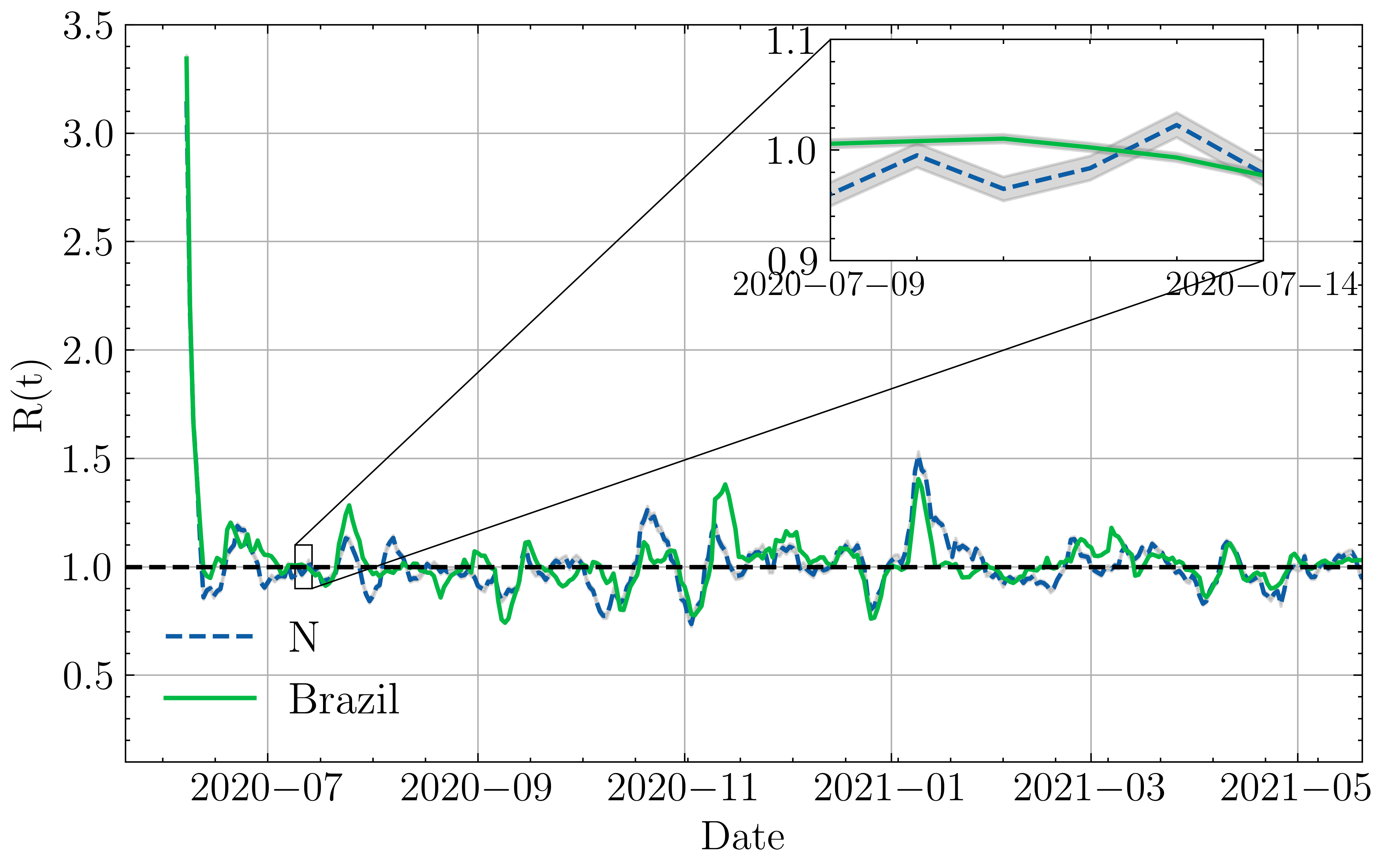}
    \caption{\footnotesize{$R(t)$ evolution in North region (N). Lines indicate the estimated mean of $R(t)$ and the shaded area delimits the $95\%$ credible interval.}}
    \label{RT_N}
\end{figure}

Brazil is divided into five major geographic regions: North (N), Northeast (NE), South (S), Southeast (SE), and Center West (CO). In Fig.~\ref{RT_SE} - \ref{RT_N}, the estimated mean of $R(t)$ as a function of the calendar time is shown for each major Brazilian geographic region and for the country as a whole. The results show a similar degree of correlation as seen in Fig.~\ref{RT_AM}-\ref{RT_MT}. The S region is the one with the highest divergence from the national curve. The similarities among the estimated $R(t)$ curves indicate that the national trends find close correspondence with regional trends. 

\section{Conclusions}
In this work, the evolution of the effective reproduction number as a function of the calendar time was analyzed for the COVID-19 pandemic in Brazil. Different scenarios with national and sub-national data on reported infections were considered. For the time interval analyzed, higher correlation levels between national and subnational data were observed for $R(t)$ estimates when compared to the moving average of reported infections. Hence, these results suggest that the $R(t)$ is a more consistent metric to monitor the rates of disease transmission across distinct national levels than the moving average of the number of reported infections. In terms of NPI policies, the results indicate that epidemiological surveillance at the national level is likely to be more effective for large population states, since they present high levels of correlation with the national trends. In contrast, different levels of correlation and credible intervals were found for states with smaller populations, indicating that those are likely to benefit more by establishing policies based on their local data. Finally, the use of more advanced epidemiological models that include e.g. sub-national population dynamics could be beneficial towards obtaining more robust estimators.

\section*{Acknowledgment}
The authors thank the Programa de Pós-Graduação em Engenharia Elétrica (PPGEE) and Departamento de Engenharia Elétrica (DEE) both from the Universidade Federal de Campina Grande (UFCG) for providing the research infrastructure used in the course of this investigation. This study was financed in part by the Coordenação de Aperfeiçoamento de Pessoal de Nível Superior - Brasil (CAPES) - Finance Code 001. 
%
\bibliographystyle{chicago}
\bibliography{Bibliography}

\end{document}